# Possible complete miscibility of $(BN)_x(C_2)_{1-x}$ alloys

Jin-Cheng Zheng[a)*], Hui-Qiong Wang [b)], A. T. S. Wee, and C. H. A. Huan
Department of Physics, National University of Singapore, Lower Kent Ridge Road, Singapore 119260
* **Corresponding author.** Email: jincheng_zheng@yahoo.com  (J-C Zheng)

**Abstract**
The stabilities of $(BN)_x(C_2)_{1-x}$ alloys and related superlattices are investigated by *ab initio* pseudopotential calculations. We find that the $(BN)_1/(C_2)_1$ superlattices in (111) orientations have the lowest formation energy among many short-range ordered $BNC_2$ structures due to the smallest number of B-C and C-N bonds. Based on the calculated formation energies at several compositions and for various ordered structures and assuming thermodynamic equilibrium, the solid solution phase diagram of $(BN)_x(C_2)_{1-x}$ alloys is constructed. We find that the complete miscibility of $(BN)_x(C_2)_{1-x}$ alloys is possible, which is in contrast with previous theoretical predictions but in agreement with experimental reports.

There has recently been considerable interest in new B-C-N ternary compounds that have potential uses as superhard materials and electrical materials. Since diamond and cubic BN are well-known materials with the highest hardness, cubic BCN compounds are expected to exhibit excellent mechanical properties and high hardness. Such cubic ternary B-C-N systems are also expected to be thermally and chemically more stable than diamond. They should also be wide band gap semiconductors since diamond and *c*-BN have band gaps of 5.5 eV and 6.1 eV respectively.

Although there have been a few reports on the synthesis of cubic B-C-N (c-BCN) materials, several outstanding issues remain. For example, Badzian[1] reported the synthesis of $(BN)_x(C_2)_{1-x}$ solid solutions, where $0.15<x_{BN}<0.6$. Knittle et al[2] reported the synthesis of cubic $BC_xN$ with x=0.9-3.0 under high pressure (30 GPa) and high temperature (laser heating to ~2000-2500K) conditions. Komatsu et al[3] synthesized c-$BC_{2.5}$N from graphitic $BC_{2.5}$N. However, limited mutual solubility of *c*-BN and diamond in the solid state was predicted by theory based on a simple pseudobinary "regular-solution" model for short-range order, where the energies of formation of the representative ordered compounds were calculated from first principles[4]. Recent first principles calculations[5-7] also showed a relatively high formation energy of cubic $BC_2N$ (alloy or superlattice (SL) of $(BN)_n/(C_2)_n$) as compared with the average cohesive energy of the parent materials, diamond and *c*-BN.

A possible reason for the disagreement between theoretical and experimental results may be due to the absence of lower-energy short-range-order structures in the theoretical models. We note that many previous calculations of cubic $(BN)_x(C_2)_{1-x}$ alloys or $(BN)_n/(C_2)_n$ superlattices were done using (001)[5,6] or (110)[4] orientations. In contrast, our present calculations show that the formation energies of $(BN)_n/(C_2)_n$ (for n < 4) for the (111) orientation are significantly smaller than for (001) or (110) orientations. In particular, for the short-range-order structure of $(BN)_1/(C2)_1$ (1+1 superlattice), the formation energy along the (111) orientation is half that along (001)

---
a) Current address: TCM, Cavendish Laboratory, Madingley Road, Cambridge CB3 0HE, UK
b) Current address: Department of Applied Physics, Yale University, New Haven, CT 06520, USA.

or (110) orientations, and it is the lowest-energy structure among many possible short-range ordered BC2N structures. The miscibility phase diagram of $(BN)_x(C_2)_{1-x}$ alloy is constructed and the complete miscibility of this system is found to be possible. The experimental realization[8] of the formation of BN/diamond (111) interface supports our total energy calculations and our model can therefore explain the experimental solid solubility of $(BN)_x(C_2)_{1-x}$ system.

Calculations of $(BN)_n/(C_2)_n$ (001), (110) and (111) superlattices and $(BN)_x(C_2)_{1-x}$ alloys are performed using the plane wave *ab initio* pseudopotential method within the local density functional (LDF) theory. Non-local norm-conserving pseudopotentials are created according to the prescription of Hamann *et al.*[9] The Hedin-Lundqvist (HL)[10] form of the exchange-correlation potential in the LDA and a mixed basis representation[11,12] is employed. The formation energy (eV/atom) of $(BN)_n/(C_2)_n$ superlattices is expressed as[14]

$$E_{form} \text{ (per atom)} = [E_{tot}^{(BN)_n/(C_2)_n} - (E_{tot}^{BN} + E_{tot}^{C_2})*n]/4n \tag{1}$$

where $E_{tot}^{(BN)_n/(C_2)_n}$ is the total energy for the $(BN)_n/(C_2)_n$ superlattice, $E_{tot}^{BN}$ and $E_{tot}^{C_2}$ are the total energies for BN and diamond respectively. The formation energy (excess enthalpy) for a specific ordered structure $s$ of $(BN)_x(C_2)_{1-x}$ alloy is defined as[5,16]

$$\Delta H(x,s) = E_{tot}^s - xE_{tot}^{BN} - (1-x)E_{tot}^{C_2} \tag{2}$$

where $E_{tot}^s$ is the total energy per atom for the alloy structure $s$. The lattice constants obtained from the present ab initio pseudopotential calculations are 3.561 Å for diamond and 3.592 Å for BN with zinc-blende structure, which are in agreement with experimental data of 3.567 Å for diamond and 3.617 Å for BN[13].

The formation energies (eV/atom)[14] of $(BN)_n/(C_2)_n$ (001), (110) and (111) superlattices as a function of thickness $n$ are plotted in Fig 1, which exhibits several interesting trends. First, all the formation energies of the $(BN)_n/(C_2)_n$ superlattices are positive, which indicates the thermodynamic instability with respect to segregation. The formation energy of $(BN)_1/(C_2)_1$ (001) is 0.432 eV/atom in this work, which is in good agreement with a previous pseudopotential result[7] of 0.433 eV/atom. Second, we find that for different orientations, the formation energies all decrease with increase of superlattice thickness, indicating the preference for phase segregation. As the superlattice thickness increases, more and more bulk-like layers are contained, leading to a decrease in the formation energy. Third, polar (111) superlattices show the lowest formation energies among all low-index $(BN)_n/(C_2)_n$ superlattices. For instance, the formation energy of $(BN)_1/(C_2)_1$ (111) is 0.210 eV/atom for the unrelaxed structure, and 0.196 eV/atom after relaxation, much lower than those of $(BN)_1/(C_2)_1$ (001) and (110) superlattices. In particular, the formation energy of $(BN)_n/(C_2)_n$ (111) is about half that of $(BN)_n/(C_2)_n$ (001), consistent with the fact that the number of unfavorable B-C or N-C bonds at the BN/C2 (111) interface is only half of that at the (001) interface. Further calculations of formation energy of $(BC)_1/(NC)_1$ (001) and (111) give 0.888 eV/atom and 1.103 eV/atom respectively for the fully relaxed geometries. These values are both much larger than those of the unrelaxed $(BN)_1/(C_2)_1$ (001) and (111) superlattices and indicate the unfavorable bonding of stacking types involving only B-C & N-C bonds. Experimental observations[8] of the preferred formation of BN/C$_2$ interface in (111) orientation support our analysis. Thus this work completes the previous studies that compute only the formation energies of polar (001)[5,6] and nonpolar (110)[4] superlattices.

In many cases the stability of ternary semiconductors is strongly related to their corresponding superlattices, since some representative short-range order structures have superlattice structures. For example, the CuPt structure[20] is a (1+1) superlattice in (111) orientation, the CuAu structure (also $L1_0$ structure)[4,5,20] is a (1+1) superlattice in (001) or (110) orientation, etc. The stability of $(BN)_x(C_2)_{1-x}$ is therefore related to the stability of $(BN)_n/(C_2)_n$ (001), (110) and (111) superlattices. The formation energies for several short-range ordered structures with various compositions are shown in Fig 2. The Connolly-Williams (CW) structures[4,5,15] include bulk diamond (x=0), L1 (Luzonite, x=0.25)[4,5], CuAu (CA, x=0.5)[4,5], L2 (Luzonite, x=0.75)[4,5], and bulk BN (x=1). The formation energies of $(BN)_x(C_2)_{1-x}$ in CW structures are largest compared with other possible structures considered in this study. Properties such as formation energy[4,5,15,16], band gap[4,16] and band offset[16-18] of disordered general $(A)_x(C)_{1-x}$ alloys (composed of A and C compounds or elements) can be obtained by atomic cluster expansion of CW structures[15]. Using CW structures as a basis set to calculate the stability of $(BN)_x(C_2)_{1-x}$ alloys, Lambretch et al[4] estimated a very high critical temperature of miscibility of $(BN)_x(C_2)_{1-x}$ alloys, even after improved by a high-temperature expansion method[19], still much higher than the melting temperature (3673 K) of $(BN)_{0.5}(C_2)_{0.5}$ alloys, which was estimated by average of melting temperature of diamond (4100 K at ~ 12.5 GPA)[13] and BN (3246 K)[13]. This theoretical prediction of very high critical temperature indicated very limited solid solution in $(BN)_x(C_2)_{1-x}$ alloys, which can't well interpret Badzian's report[1] of the synthesis of solid solutions within the compositional range $0.15<x_{BN}<0.6$. Two possible explanations could account for this discrepancy: (i) Mixing in the experiments took place in the liquid state and the solid solutions are metastable systems, as pointed out by Lambrecht et al[4]; (ii) additional short-range ordered structures should be included to form a representative basis for all zinc blende structures to account for possible short-range interactions, as shown by Ferreira et al[20]. Specifically, the (111) orientation should be included, which we have shown to be the lowest-energy possible structure (Fig. 2).

To estimate the lower limit of critical temperature, we considered additional short-range ordered structures of $(BN)_x(C_2)_{1-x}$, as follows: the Famatinite (F1 or F3) structure (related to $x$=0.25 or 0.75 in $(BN)_x(C_2)_{1-x}$); SL1+3(001), SL1+3(110), SL1+3(111) structures in $(BN)_n/(C_2)_n$ superlattices corresponding to $x$=0.25; SL1+1(001), SL1+1(110), SL1+1(111) structures corresponding to $x$=0.5; and SL3+1(001), SL3+1(110), SL3+1(111) structures corresponding to $x$=0.75$_{xx}$. We also calculate the formation energies of all possible cubic BC2N structures in an eight-atom zinc-blende cubic unit cell, as proposed by Sun et al[7]. The lowest formation energies of such structures are comparable to that of SL1+1(001), and much higher than that of SL1+1(111). We did not include other high-formation-energy structures in the phase diagram. From Fig 2, it can be seen that the formation energies of the short-range ordered structures with the same composition follow the relationship:
$\Delta H(SL(111)) < \Delta H(SL(110)) < \Delta H(SL(001)) < \Delta H(Famatinite) < \Delta H(Luzonite)$
The superlattice structures are more stable than ordered structures such as the Famatinite and Luzonite structures in $(BN)_x(C_2)_{1-x}$ alloys. This result is in contrast to that of $GaAs_{1-x}N_x$ alloys[21], where $\Delta H(Famatinite) < \Delta H(SL(111)) < \Delta H(Luzonite)$. The low formation energies of SL(111) further indicate the phase separation behaviors (from mixed alloy to $BN/C_2$ superlattices) in $(BN)_x(C_2)_{1-x}$ alloys.

In our work, the solubility limit in $(BN)_x(C_2)_{1-x}$ alloys is estimated based on two main assumptions: (a) thermodynamic equilibrium and (b) the calculated formation energies being a lower limit for all possible structures, as adopted by Neugebauer and Van de Walle for $GaAs_{1-x}N_x$ alloys[21]. We use the calculated formation energy to construct a lower limit $\Delta H^{min}(x)$ for each composition[21],

$$\Delta H^{min}(x) = 4x(1-x)\Delta H^0 \quad (3)$$

where $\Delta H^0 = 0.1528$ eV ($\Delta H^{min}(x) = 0.1146$ with x=0.75 for SL 3+1 (111)). The miscibility gap is analytically estimated and its behavior as a function of temperature is given by the binodal line, as shown in Fig 3 (solid line)[4,21].

$$k_B T/\Delta H^0 = (8x-4)/[\ln x - \ln(1-x)] \quad (4)$$

The region below the spinodal line is unstable, as indicated by dash-dot line[4,21]

$$k_B T/\Delta H^0 = 8x(1-x) \quad (5)$$

The critical temperature of the miscibility gap is thus $T_{GM} = 2\Delta H^0/k \approx 3546$ K, which is lower than the melting temperature (3673 K) of $(BN)_{0.5}(C_2)_{0.5}$ alloys. Hence, with this approach, the complete miscibility of $(BN)_x(C_2)_{1-x}$ alloys is possible and interprets experimental reports[1-3].

In conclusion, we have used *ab initio* pseudopotentials calculations to investigate the stabilities of $(BN)_n/(C_2)_n$ (001), (110) and (111) superlattices and $(BN)_x(C_2)_{1-x}$ alloys. $(BN)_n/(C_2)_n$ superlattices in (111) orientation are found to have the lowest formation energy compared to (110) and (001) orientations. These results suggest that in the growth of diamond on BN or BN on diamond or mixed BCN alloys, it is more favorable to form BN/diamond interfaces or superlattices in the (111) rather than (110) or (001) orientations. Based on the calculated formation energies at several compositions and for various ordered structures, as well as assuming thermodynamic equilibrium, the complete miscibility of $(BN)_x(C_2)_{1-x}$ alloys is shown to be possible.

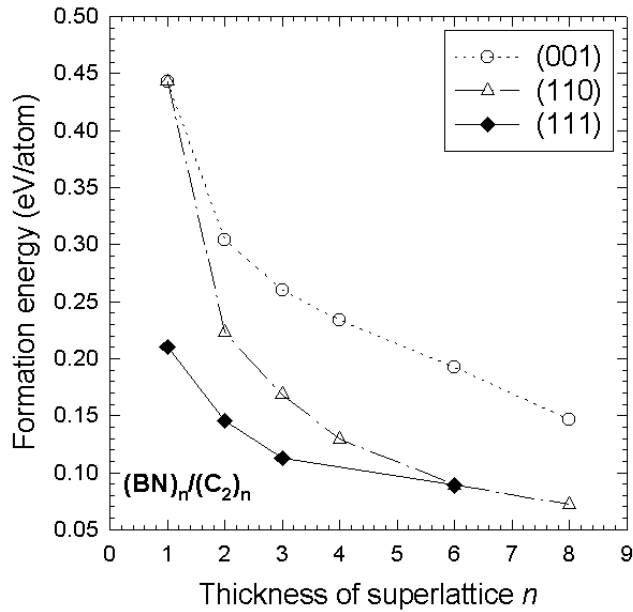

Fig 1. The formation energy (eV/atom) of $(BN)_n/(C_2)_n$ (001), (110) and (111) superlattices as a function of thickness $n$.

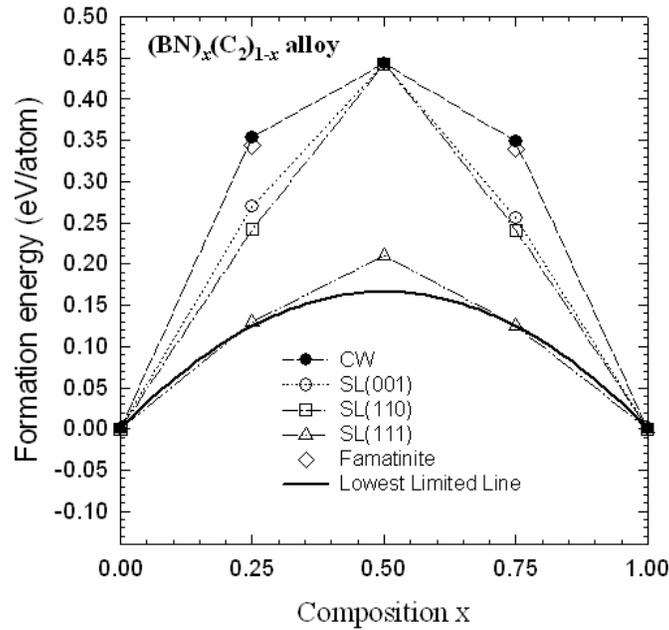

Fig 2. The formation energy (eV/atom) for several compositions and for various ordered structures of $(BN)_x(C_2)_{1-x}$ alloys.

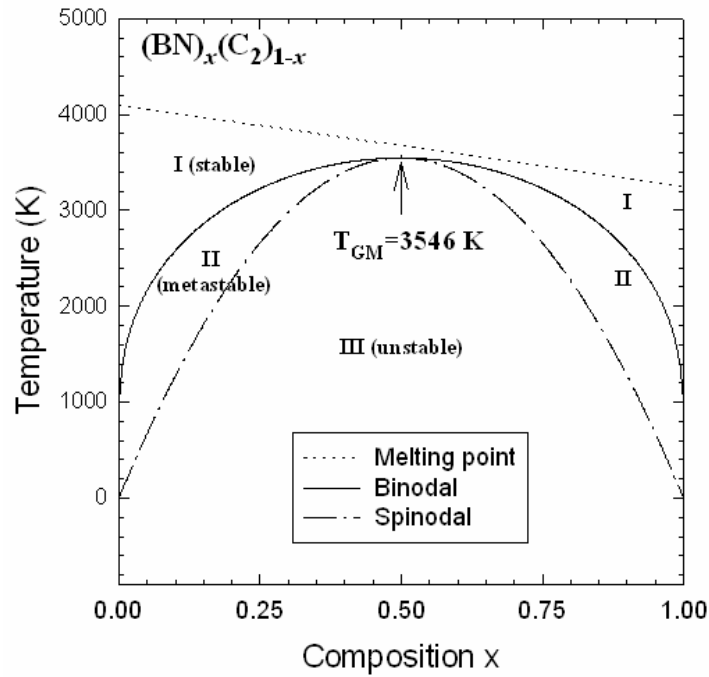

Fig 3. The phase diagram of $(BN)_x(C_2)_{1-x}$ alloys with lower limit of the miscibility gap. The binodal line is the solid curve, and spinodal line is the dash-dot curve. The liquidus line which separates the liquid from the solid phase is obtained by linearly averaging the melting points of diamond and cubic BN (dotted curve).

6